\documentclass[twocolumn]{aastex62}

\usepackage{amsmath}
\usepackage{cases}
\usepackage{color}

\newcommand{\beq}{\begin{equation}}
\newcommand{\eeq}{\end{equation}}

\newcommand{\hi}{H{\sc i}~}
\newcommand{\HI}{H{\sc i}}
\newcommand{\kms}{km ${\rm s^{-1}}$~}
\newcommand{\kmsa}{km ${\rm s^{-1}}$}

\newcommand{\nhi}{$\mathrm{N}_\mathrm{H{\sc I}}$}
\newcommand{\nHI}{$\mathrm{N}_\mathrm{H{\sc I}}~$}

\newcommand{\Ms}{$\mathcal{M}_s$}

\received{February 4, 2019}
\revised{February 25, 2019}
\accepted{February 26, 2019}

\shorttitle{\hi intensity structure}
\shortauthors{Clark et al.}

\begin{document}

\title{The Physical Nature of Neutral Hydrogen Intensity Structure}

\author[0000-0002-7633-3376]{S.E. Clark}
\altaffiliation{Hubble Fellow}
\affiliation{Institute for Advanced Study, 1 Einstein Drive, Princeton, NJ 08540, USA}

\author[0000-0003-4797-7030]{J.E.G. Peek}
\affiliation{Space Telescope Science Institute, 3700 San Martin Dr, Baltimore, MD 21218, USA}
\affiliation{Department of Physics \& Astronomy, Johns Hopkins University, Baltimore, MD 21218, USA}

\author[0000-0002-7351-6062]{M.-A. Miville-Desch\^enes}
\affiliation{AIM, CEA, CNRS, Universit\'e Paris-Saclay, Universit\'e Paris Diderot, Sorbonne Paris Cit\'e, F-91191 Gif-sur-Yvette, France}

\correspondingauthor{S.E. Clark}
\email{seclark@ias.edu}

\begin{abstract}
We investigate the physical properties of structures seen in channel map observations of 21-cm neutral hydrogen (\HI) emission. \hi intensity maps display prominent linear structures that are well aligned with the ambient magnetic field in the diffuse interstellar medium (ISM). Some literature hold that these structures are ``velocity caustics", fluctuations imprinted by the turbulent velocity field, and are not three-dimensional density structures in the ISM. We test this hypothesis by stacking probes of the density field -- broadband far infrared (FIR) emission and the integrated \hi column density (\nhi) -- at the locations of linear \hi intensity structures. We find that the \hi intensity features are real density structures and not velocity caustics. We generalize the investigation to all small-scale structure in \hi channel maps, and analyze this correlation as a function of velocity channel width, finding no measurable contribution from velocity caustics to the \hi channel map emission. Further, we find that small-scale \hi channel maps structures have elevated FIR/\nhi, implying that this emission originates from a colder, denser phase of the ISM than the surrounding material. The data are consistent with a multi-phase diffuse ISM in which small-scale structures in narrow \hi channel maps are preferentially cold neutral medium (CNM) that is anisotropically distributed and aligned with the local magnetic field. The shallow spatial power spectrum (SPS) of narrow \hi channels is often attributed to velocity caustics. We conjecture instead that the small-scale structure and narrow linewidths typical of CNM explain the observed relationship between the SPS and channel width.
\end{abstract}


\section{Introduction}
The diffuse interstellar medium (ISM) fills most of the volume of the Milky Way. This material, the progenitor of dense structures that eventually form stars, is multi-phase, magnetized, and turbulent. The ISM is broadly partitioned into ionized, molecular, and atomic components, with the atomic gas composed of a cold neutral medium (CNM), warm neutral medium (WNM), and a thermally unstable component \citep[for reviews, see][]{Ferriere:2001, Cox:2005}. Understanding the distribution of matter and energy within and between these phases is a major goal of ISM research. Since its initial detection in 1951 \citep{Ewen:1951, Muller:1951}, the 21-cm line from the hyperfine transition of neutral hydrogen (\HI) has been an extremely fruitful probe of the distribution of interstellar gas \citep[e.g.][]{Kalberla:2009}. 

The ISM occupies three spatial dimensions, but our direct observations are limited to the two-dimensional plane of the sky. A third dimension is available to spectroscopic observations, which record intensity as a function of frequency. Emission line observations like 21-cm \hi measurements are typically reported in position-position-velocity (PPV) space, where the third dimension represents the line-of-sight velocity implied by the frequency shift from the line rest frequency. Such shifts are not caused exclusively by bulk gas motions, but also by thermal and turbulent line broadening. This PPV information is therefore distinct from position-position-position (PPP) space, but is also not merely a projection of the six-dimensional spatio-kinematic information that describes flows of interstellar material. This observational restriction complicates the interpretation of ISM structures, particularly the determination of how observed PPV intensity structures relate to ``real" PPP density structures \citep{Beaumont:2013}. This is important for understanding molecular clouds, which are commonly probed by PPV molecular line observations \citep{Goodman:2011}. 

In simple flows, there exists a one-to-one mapping between PPP and PPV, e.g. in a pure Hubble flow. At least in theory, it is possible to construct a circumstance in which the structure seen in PPV space can be dominated by the velocity field, and thus not a simply-interpretable probe of the density field. This can be understood by considering the extreme case: incompressible, constant-density turbulence. Such an idealized flow will lack any PPP density structure, but can still exhibit PPV intensity structure. A featureless incompressible flow is far from a realistic description of the ISM \citep[e.g.][]{Elmegreen:2004}. Still, it is reasonable to interrogate to what extent intensity structures in \hi PPV observations are influenced by the true density or velocity fields. 

Recent high-dynamic range observations of the diffuse ISM have renewed interest in this question. \hi surveys such as the Galactic Arecibo $L$-Band Feed Array Survey \citep[GALFA-\HI;][]{Peek:2018} and \HI4PI \citep{HI4PI:2016} show a wealth of linear structure, particularly prominent at high Galactic latitudes and in narrow ($\sim$few \kmsa) velocity channels. \citet{Clark:2014} developed a machine vision algorithm called the Rolling Hough Transform (RHT) to measure the linearity of image structure, and used the RHT to show that \hi intensity structure is well aligned with the ambient magnetic field as traced by optical starlight polarization. With the advent of all-sky maps of polarized dust emission at 353 GHz \citep{PlanckXIX:2015}, \citet{Clark:2015} measured a robust, tight correlation between linear \hi intensity structures and the magnetic field. Similar structures have been seen in other \hi emission data observed at high angular resolution \citep{Martin:2015,Kalberla:2016, Kalberla:2017, Blagrave:2017}. These \hi intensity observations are reminiscent of linear, magnetically aligned structures in other media, including \hi absorption \citep{McClure-Griffiths:2006} and Faraday depth structures in low-frequency radio polarimetric data \citep{Jelic:2015}. 

Do these results indicate that the true (PPP) density field is striated, and that these striations are preferentially elongated along the ambient magnetic field? Or is this an effect of the velocity field? A number of explanations have been proposed for the origin of this magnetically aligned structure. Most theoretical explanations have focused on physical mechanisms for a correlation between the orientation of elongated density structures and the magnetic field \citep[e.g.][]{Hennebelle:2013tr, Inoue:2016, Soler:2017}. 

However, the striking magnetic alignment of \hi structures in velocity channel maps \citep{Clark:2014, Clark:2015} could in theory be an imprint of the velocity field. This interpretation has been put forward by \citet{LazarianYuen:2018}, based on the theory of \citet{LP2000}: that the structures seen in these \hi channel maps are ``velocity caustics": intensity structures created by turbulent velocity fluctuations coincident at a given $v_{lsr}$. In this picture, the measured alignment between the magnetic field orientation and the orientation of \hi intensity structures has little or no relation to the underlying density field, but is instead caused by the correlation between velocity and magnetic field orientation that is predicted from magnetohydrodynamic (MHD) turbulence \citep{Goldreich:1995, Elmegreen:2004}. Shearing turbulent eddies that are elongated in the direction of the local magnetic field have a velocity gradient that is steepest in the direction perpendicular to the magnetic field. If \hi intensity structures are dominated by velocity caustics, the correlation between the gradient of the intensity in thin velocity channel maps and the dust polarization angle would be a measurement of the velocity-magnetic field alignment \citep{LazarianYuen:2018}.

In this work, we apply a direct test to determine whether the linear structures identified in narrow spectral channels are velocity caustics or real density structures. We measure the correlation between linear channel map \hi structures and the density field. We further generalize our investigation to all small-scale structure in thin-channel \hi intensity maps. This paper is structured as follows. In Section \ref{sec:data} we detail the data products and numerical simulations used in this work. In Section \ref{sec:physnature} we test whether the structure seen in \hi intensity maps is consistent with being velocity caustics, and investigate the physical nature of structures observed in \hi intensity. In Section \ref{sec:discussion} we discuss how our findings fit into a broader picture of the neutral ISM, and suggest several testable predictions. In Section \ref{sec:conclusions} we summarize and conclude.

\begin{figure*}
\centering
\includegraphics[width=\textwidth]{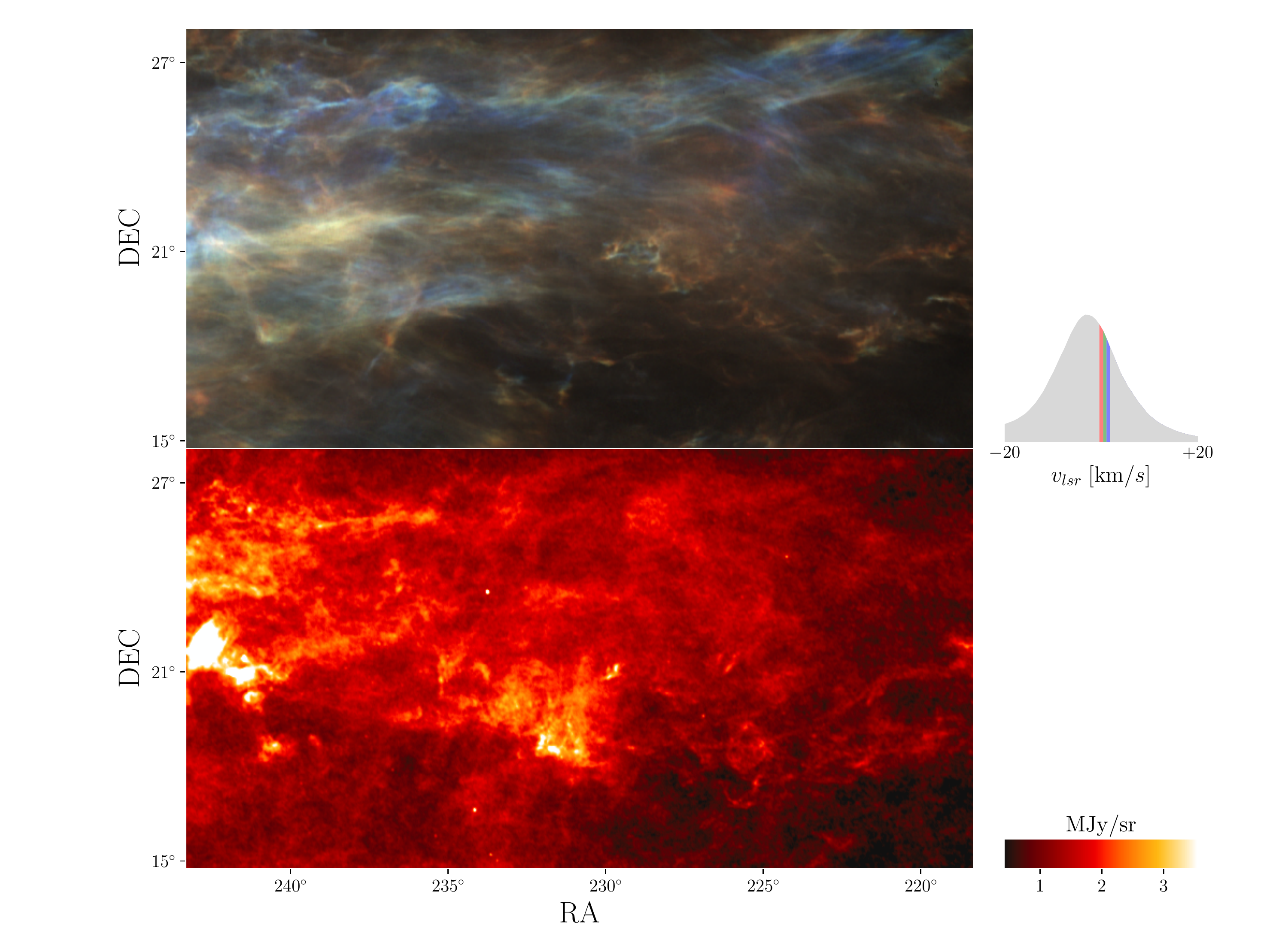}
\caption{A portion of the data analyzed. Top: three-color image of three adjacent \hi velocity channels where the filamentary structure quantified in \citet{Clark:2014} is visually evident. Side panel shows the integrated \hi brightness temperature spectrum for the region described in Section \ref{sec:data}. The highlighted velocity channels are shown in red, green, and blue in the image. Each velocity channel has a width $\delta v = 0.72$ \kmsa. Bottom: \textit{Planck} 857 GHz emission in the same region. Several FIR point sources are visible in the $I_{857}$ map: these are masked from our analysis according to the \textit{Planck} HFI point source mask.   
}\label{fig:RGB_FIR}
\end{figure*}

\section{Data and Simulations}\label{sec:data}

In this work we analyze 21-cm line observations of Galactic \hi and measurements of the far infrared (FIR) intensity at 857 GHz.
We make use of several publicly available \textit{Planck} legacy data products. Much of this work uses the 857 GHz frequency map from the \textit{Planck} collaboration data release R3.01 \citep{Planck2018I}. We also use the High Frequency Instrument (HFI) point source mask to exclude regions of the data that are identified as point sources in the FIR \citep{PlanckXXVI:2016}. The zero-point of the \textit{Planck} 857 GHz emission is significantly uncertain, and the data must be corrected for the cosmic infrared background monopole. In this work we apply a monopole offset correction of $0.64$ MJy/sr \citep{PlanckVIII:2016}. However, we intentionally design the experiments in this paper such that our results are not qualitatively dependent on this value.

\begin{figure*}
\centering
\includegraphics[width=0.9\textwidth]{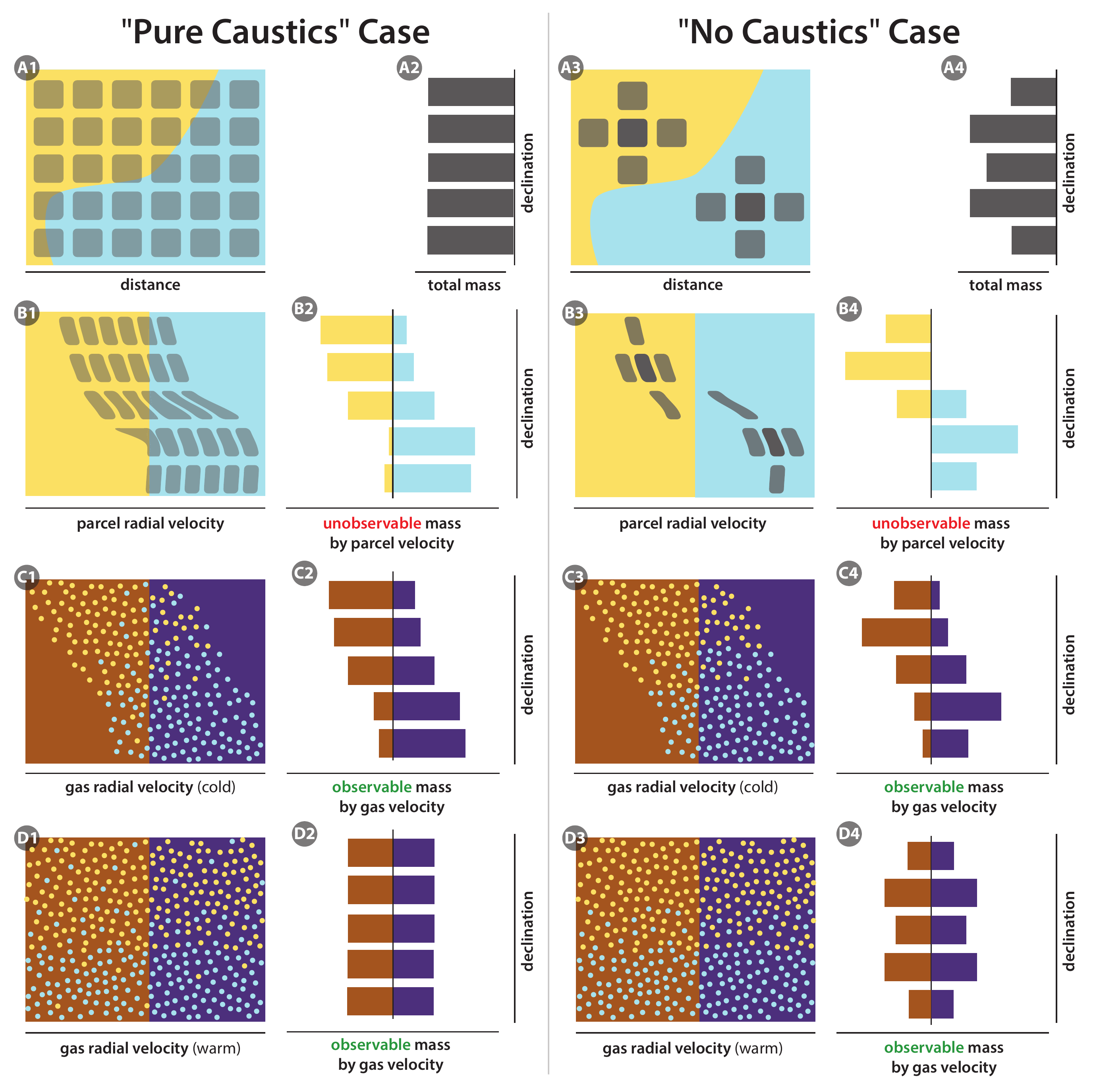}
\caption{Illustration of the relationship between the density and velocity of interstellar gas and observable quantities. At issue is the relationship between structure along the line of sight and spatial structure on the plane of the sky. Without loss of generality we depict spatial structure along a single dimension (declination). A1, A3: Gray squares represent the three-dimensional matter distribution in declination and distance from the observer. Yellow and blue regions represent different line-of-sight velocities, such that the boundary between them represents an isovelocity contour amid a velocity gradient. A2, A4: Column density map, or total mass as a function of declination. B: Parcels of mass mapped to their radial velocity. B1 and B3 show the mass distribution as a function of radial velocity: this is simply A1 and A3 in a different coordinate system. B2, B4: mass tabulated by parcel velocity. B2 and B4 are unobservable because of the finite width of emission lines. C and D show observable signatures of these velocity and density fields for different gas temperatures. Each panel in C and D shows two observational velocity channels in brown and purple. C1, C3, D1, D3: sample gas distributions, color coded by their parcel radial velocity from A and B. Thermal broadening causes the distribution of gas radial velocities to be wider than the parcel radial velocities in B. C2, C4, D2, D4: observable mass by gas velocity, or measured gas emission as a function of declination in two channel maps.
Left panel depicts the extreme case of ``pure caustics''. There is no density variation on the scale of the velocity gradient (A1) and the column density is constant as a function of position on the sky (A2). If this gas is cold enough that its thermal linewidth is comparable to the observed velocity channel width, it is possible to see velocity caustics: structure in the channel maps induced purely by the velocity field (C2). This structure will be less severe than the unobservable case without thermal broadening (B2) and the velocity caustic structure will be washed out entirely if the thermal linewidth is large compared to the scale of the velocity gradient (D1). In this case the channel map structure will be strongly correlated with the column density map (D2). 
Right panel shows a case with no velocity caustics, where structure in the column density map (A4) is entirely a function of true three-dimensional density variation (A3). If this gas is cold, the channel map will show structure that maps to true density variation along the plane of sky and distance spatial dimensions (C4). If this gas is warm, the channel map images will again be highly correlated with the column density map (D4). 
}\label{fig:causticscartoon}
\end{figure*}

We use the publicly available Data Release 2 (DR2) of GALFA-\hi \citep{Peek:2018}. GALFA-\hi is a large-area ($\sim4$ sr) survey with high spatial ($4'$) and spectral ($0.18$ \kmsa) resolution. We use three of the data products provided in DR2: PPV cubes of the \hi brightness temperature ($T_b(v)$), the RHT output of GALFA-\hi velocity channel map data, and the stray radiation-corrected column density map ($\mathrm{N}_\mathrm{H{\sc I}}$), derived from $T_b(v)$ integrated over $\left|v\right| < 90$ \kmsa. While \hi is susceptible to self absorption effects in very cold, dense media, this phase is relatively uncommon in the low-density ISM under investigation here \citep{MurrayPeek:2018}. In this work we ignore this effect and treat velocity-integrated brightness temperature and column density interchangeably. In what follows, an \hi ``channel map" refers to GALFA-\hi $T_b(v)$ integrated over a narrow range in velocity:

\beq\label{eq:IHI}
I_{\mathrm{H{\sc I}}}(\vec{x}) = \int_{v_1}^{v_2} T_b (\vec{x}, v) \delta v,
\eeq

where $\vec{x}$ represents the plane-of-sky spatial dimensions. The velocity ``channel width" is $\delta v = v_2 - v_1$, and the channel is centered on $v_0 = (v1 + v2)/2$. \nHI is calculated via Equation \ref{eq:IHI} with $v_1 = -90$ \kms and $v_2 = +90$ \kmsa, and reported in units of $\mathrm{cm}^{-2}$.

Unless otherwise noted, the analysis in this work is carried out on a large region of the GALFA-\hi sky. We select data in the GALFA-\hi footprint at Galactic latitudes $|b| > 30^\circ$. We exclude sightlines in the HFI point source mask, as well as those with \nHI $ > 8\times 10^{20}~\mathrm{cm}^{-2}$, to focus on the diffuse ISM. The resulting sky area considered comprises $\sim 50\%$ of the GALFA-\hi sky, or $\sim 2$ sr. A portion of the \textit{Planck} 857 GHz emission and the GALFA-\hi channel map data is shown in Figure \ref{fig:RGB_FIR}.

Finally, to illustrate the relationship between the channel map structure and the column density structure of turbulent flows, we run a series of simple isothermal simulations. These are three-dimensional hydrodynamic simulations produced with RAMSES \citep{Teyssier:2002}. We run simulations at three sonic Mach numbers (\Ms=0.5, 1, 5), each with size 128$^3$, and translate these to PPV space with a velocity resolution $\delta v_0 = 0.1$ \kmsa. 

\section{The physical nature of thin-channel \hi intensity structures}\label{sec:physnature}

Broadband observations of thermal dust emission are sensitive to the density, but not velocity, field. No velocity caustics exist in dust emission maps. The observed correlation between the orientation of filamentary structures in dust emission and the dust polarization field \citep{PlanckXXXII:2016, PlanckXXXV:2016, Malinen:2016, SolerVelaC:2017} is thus unequivocally a correlation between density structures and the projected magnetic field. \nHI is likewise sensitive to the density field, but not the velocity field, because of the integration over a broad velocity range (see Section \ref{sec:data}). 

\subsection{Theoretical considerations}

Before undertaking any data analysis, we ask under what physical conditions we might expect the thin-slice \hi intensity to be dominated by velocity caustics. Here we should emphasize an important distinction between the velocity caustics picture and other questions of PPV structure. Velocity caustics occur when the dominant intensity contours -- the spatial structures -- in a velocity channel are imprinted by the turbulent velocity field rather than the density field. The absence of velocity caustics does not imply that the distribution of emission along the line of sight is unaffected by the radial velocity. The classic example of an expanding shell, for instance, will cause emission from a single PPP structure to contribute to multiple line-of-sight velocities. This type of velocity-induced emission structure, and its inverse, whereby emission at one velocity originates from multiple density structures along the line of sight, certainly complicates the relationship between PPP and PPV space \citep[see Figure 1 of][]{Beaumont:2013}. The difficulty of reverse-engineering PPP density structures from PPV observations is a long-standing problem \citep[e.g.][]{Ostriker:2001, Clarke:2018}. 
The velocity caustics picture tested here is a distinct and specific claim: that the intensity structure in narrow velocity channels is an image of the turbulent velocity field. Put another way, the intensity structure observed at line-of-sight velocity $v_0$ converges to the constant-density turbulent velocity field at $v_0$ as the channel width $\delta v$ becomes small enough. This turbulent velocity field can be uncorrelated with the density field, as explicitly assumed in the \citet{LP2000} treatment of incompressible turbulence. 

\begin{figure*}
\centering
\includegraphics[width=\textwidth, trim=2.5cm 1cm 2.5cm 0cm]{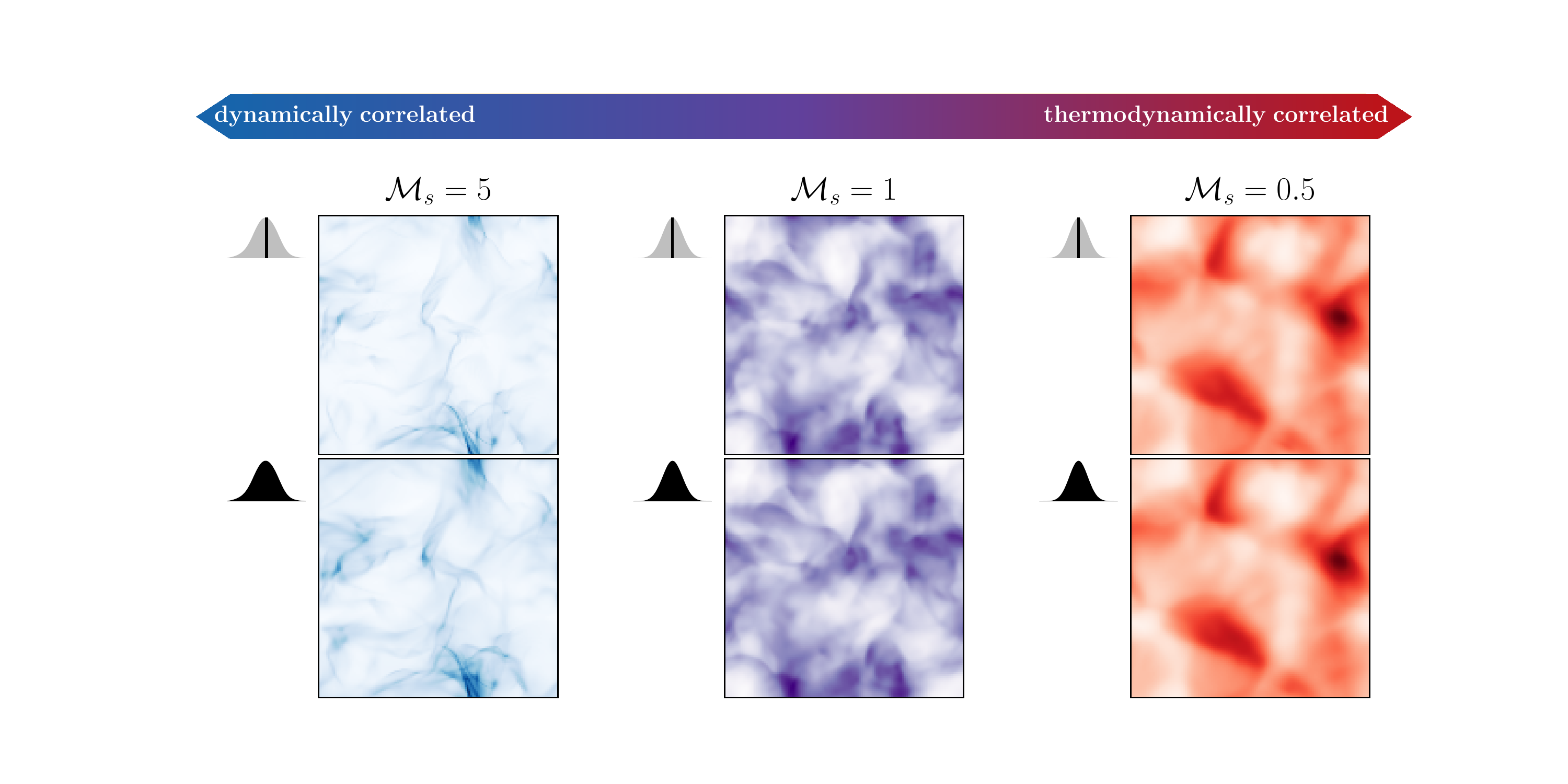}
\caption{A demonstration of the correlation between velocity channel structure and column density structure for turbulent flows with different sonic Mach numbers. From left to right, we show snapshots of simulations with \Ms$=5, 1, 0.5$. In all cases the top panel shows a single velocity channel, $\delta v = 0.1$ \kmsa, through the center of the synthetic PPV cube. The bottom panel shows the synthetic column density map: the integrated intensity over the full simulation domain. To the left of each panel we show the intensity as a function of velocity integrated over the two spatial dimensions of the simulation, and highlight the portion of the spectrum shown. In all cases the channel map structure near \hi line peak is well correlated with the column density structure, but the physical reason for this changes as a function of \Ms. For supersonic flows, the density and velocity fields are dynamically correlated. For subsonic flows, the density and velocity fields are not well correlated, but thermal broadening washes out intensity fluctuations coupled to the turbulent velocity field, and thermodynamically correlates the channel map structure with the column density map.     
}\label{fig:sims}
\end{figure*}

A significant velocity caustics contribution to synthetic channel map emission has been reported in a number of simulation studies. These results are often interpreted as meaning that channel map observations of \hi contain little to no structure that can be associated with density structures. The relevance of this picture for PPV observations of \hi is partly a question of how well these simulations capture the salient ISM physics, and partly a question of whether the transformation from the simulated PPP space to mock PPV observations correctly mimics the process by which 21-cm emission maps to line-of-sight velocity. Unphysical transformations between PPP and PPV space can generate misleading effects. 

One early exploration relating the morphology of synthetic channel maps of MHD simulations to the underlying velocity and density fields was undertaken by \citet{Pichardo:2000}. The authors compared the structure of slices through the PPP velocity field along the line of sight ($v_z$) to synthetic channel map data. \citet{Pichardo:2000} concluded that their channel maps were more morphologically similar to the $v_z$ slices than to slices through the PPP density field, and that the channel maps exhibited more small-scale structure than the density slices. This result was in qualitative agreement with the contemporary work of \citet{LP2000}. 

However, as the authors note, \citet{Pichardo:2000} neglect thermal broadening in their computation of channel map structure. Translating PPP simulations into PPV space requires the inclusion of thermal line broadening, which is a convolution along the velocity axis by a Gaussian of $\sigma=\sqrt{k_B T / m}$, where $T$ is the gas temperature, $m$ is the particle mass, and $k_B$ is the Boltzmann constant. Thermal line broadening is not a radiative transfer effect, but rather a fundamental property of thermodynamics. If this temperature-dependent convolution along the velocity axis is ignored, the influence of the turbulent velocity on channel map structures is artificially inflated. The same omission of thermal broadening is made in \citet{LazarianPogosyanVSPichardo:2001}, another work that finds prominent velocity caustics in simulations and concludes that features in observational channel maps must also be velocity structures. \citet{MAMD:2003} noted the severe limitations that thermal broadening places on the feasibility of using the \citet{LP2000} method to derive properties of the turbulent velocity field from the structure of narrow channel maps.  

The velocity caustics picture and the effects of thermal broadening are illustrated in Figure \ref{fig:causticscartoon}. Velocity caustics manifest as channel map structures induced by the three-dimensional velocity field, but weakly or not at all correlated with the three-dimensional density field (Figure \ref{fig:causticscartoon}, panel C2). This will not be observable if thermal broadening washes out velocity fluctuations (panel D2) or if three-dimensional density structures dominate the emission contours (panels C4, D4). Omitting thermal broadening from synthetic observations is particularly pernicious because it implies that channel maps directly measure the mass distribution as a function of radial velocity (panels B2 and B4). This is, in reality, unobservable, and thus cannot be compared to any real data.

\begin{figure*}
\centering
\includegraphics[width=\textwidth]{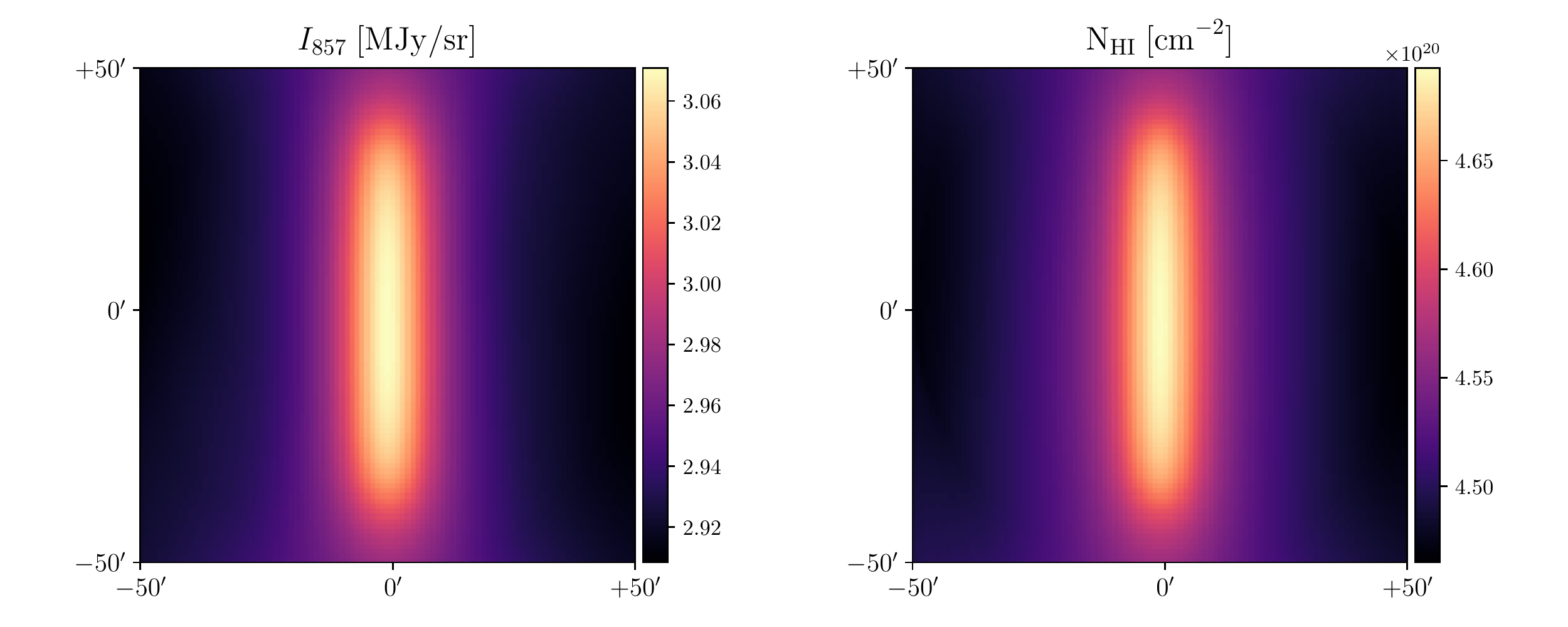}
\caption{\textit{Planck} 857 GHz intensity (left) and GALFA-\hi column density over $\left|v\right| < 90$ \kmsa (right) stacked on the RHT of GALFA-\hi channel data integrated over $v=0.3$ to $3.3$ \kmsa. Stacking on the RHT intensity $R_{v}(x, y, \theta)$ creates stacked images in every orientation bin $\theta$ on the sky. Here we rotate each stacked image by its orientation $\theta$ and average over all orientations, such that the highlighted features are all oriented along the $y$-axis of the plot. The linear features in $\delta v \sim3$ \kms channel maps highlighted by the RHT are known to be aligned with the ambient magnetic field \citep{Clark:2014, Clark:2015}. If these \hi intensity features were velocity caustics, there would be no excess of either the 857 GHz intensity or the \hi column density at the location of the features. The clear presence of correlated density structures in both data sets is inconsistent with the expectation that the \hi intensity structures are velocity caustics. 
}\label{fig:stack_857_NHI_DGR}
\end{figure*}

\citet{Burkhart:2013} examined the effect of the sonic Mach number (\Ms) on the relationship between structures in PPP and PPV. High \Ms~flows can contain shocks, which sweep up matter and thus strongly correlate the turbulent velocity field with the density field. Low $\mathcal{M}_s$ flows do not strongly correlate the velocity and density structures. However, \Ms~is the ratio of the turbulent velocity to the sound speed in the medium, which is set by the kinetic temperature. \citet{Burkhart:2013} neglected thermal broadening in much of their analysis, and so the PPV cubes they analyze are not the PPV-space representation of turbulent flows, but instead are histograms of the turbulent velocity. Assuming that these are the PPV-space representation of flows with a given \Ms~leads to the conclusion that PPP and PPV space are more closely related for higher \Ms~flows, summarized in the abstract of \citet{Burkhart:2013} as ``the dominant emission contours in PPP and PPV are related for supersonic gas but not for subsonic". 

Missing from this analysis is the fact that low \Ms~flows have higher temperatures relative to their turbulent velocities, and thus broader emission profiles. PPV cubes of low \Ms~flows that correctly incorporate the thermal line broadening have features that are smooth in velocity space and highly correlated with the PPP density structure. These lack velocity caustics: any given velocity slice of a low \Ms~PPV cube is highly correlated with the integrated line emission, especially for lightweight carriers such as \hi where this effect is strong. Thus, while PPP and PPV structures are correlated dynamically for high \Ms~flows, they are also correlated thermodynamically for low \Ms~flows. 

We demonstrate this in Figure \ref{fig:sims} using the isothermal turbulent simulations described in Section \ref{sec:data}. These are hydrodynamic simulations that are not representative of the ISM, particularly for their lack of magnetic fields and thermal structure. This is intentional, as we illustrate here a simple point about \Ms~and thermal broadening, that holds irrespective of those physics or of the particular simulation scheme used. Figure \ref{fig:sims} shows narrow channel maps of each synthetic 21-cm PPV cube for $v_0 = 0$, along with its corresponding column density map. In each simulation, the channel map structure is well correlated with the integrated intensity (column density) structure, but the reason for this correlation changes as function of \Ms. The supersonic simulation has correlated density and velocity fields that cause structures in the channel map to be well correlated with column density structures. The subsonic simulation has channel map structures that are correlated with the column density by thermal broadening. Between the dynamic and thermodynamic effects, none of the synthetic channel maps display prominent velocity caustics. 

If 21-cm PPV data contains prominent velocity caustics, then, our thermodynamic expectation is that these structures would have to be produced by the cold, supersonic gas in the diffuse \HI. As noted, the expectation from theoretical studies of turbulence is that this medium will have correlated three-dimensional density and velocity fields. One might therefore not expect that intensity channels show significant velocity-induced structure that is uncorrelated with the density field. Still, the idea that thin \hi channel maps are dominantly structured by the turbulent velocity field is pervasive in the literature. We have devised a direct observational test of this picture.

\subsection{Magnetically aligned \hi intensity structures are not velocity caustics}\label{sec:RHTnotcaustics}

We conduct an experiment to test whether thin-channel \hi intensity structures are an imprint of the velocity field. We stack FIR emission on the locations of the \hi intensity structures. As noted above, FIR emission probes the total density, but does not at all trace the velocity field. If the \hi intensity structures are velocity caustics, there should be no enhancement in the FIR emission strength at the location of the intensity structures, relative to the surrounding medium. Conversely, an enhancement in the FIR emission indicates that the structures are true density features.

To highlight the linear structures in \hi emission we use the RHT \citep[]{Clark:2014}. The RHT is a machine vision algorithm that quantifies linearity as a function of orientation, and has been used to trace the orientation of \hi emission structures that are prominently aligned with the local magnetic field \citep{Clark:2014, Clark:2015, Jelic:2018}, as well as to measure linearity in other data. The RHT output for $I_{v_0}(\vec{x})$, an \hi channel map centered at velocity $v_0$, is $R_{v_0}(\vec{x}, \theta)$, the linear intensity as a function of orientation ($\theta$) for every image pixel ($\vec{x}$). \citet{Clark:2014} referred to the narrow \hi channel map structures that are well aligned with the magnetic field as ``\hi fibers".

We measure the correlation between structures in RHT intensity and in the $I_{857}$ and \nHI maps, by stacking $I_{857}$ and \nHI on $R_{v_0}(\vec{x}, \theta)$. Here, $R_{v_0}(\vec{x}, \theta)$ is the RHT intensity for a channel map with $\delta v = 2.94$ \kmsa, centered at $v_0 \sim 1.5$ \kmsa. We stack a $101' \times 101'$ region of sky around each nonzero pixel in the RHT output. We perform a separate stack on the RHT intensity maps at each orientation angle. We weight each of these sky regions by the $R_{v_0}(\vec{x}, \theta)$ value of the central pixel. The result of this process, for each stacked map, is a $101' \times 101'$ stacked map for each orientation bin of the RHT output. We collapse the orientation information by rotating each stacked map by its orientation, such that each map is aligned with $\theta=0^\circ$. The GALFA-\hi data contains telescope scan artifacts at specific orientations because of the survey's ``meridian nodding" or ``basketweave'' scan strategy \citep[see][]{Peek:2011fp}. We try cutting orientations near $180^\circ$ that are known to be most contaminated by these artifacts, and verify that even conservative cuts do not meaningfully alter our results.

Figure \ref{fig:stack_857_NHI_DGR} shows the result of this stacking experiment. We emphasize that if the thin-channel \hi intensity structures were velocity caustics, they would have no correlation with the total density tracers, and neither of these stacks would show an enhancement at the location of the $R_{v_0}(\vec{x}, \theta)$ emission. Clearly, there is a significant enhancement both in 857 GHz emission and in \nHI at the location of the \hi intensity structures. A description of the magnetically aligned \hi fibers as velocity caustics is inconsistent with this result. 

\begin{figure*}
\centering
\includegraphics[width=\textwidth, trim=4.5cm 1cm 3cm 0cm]{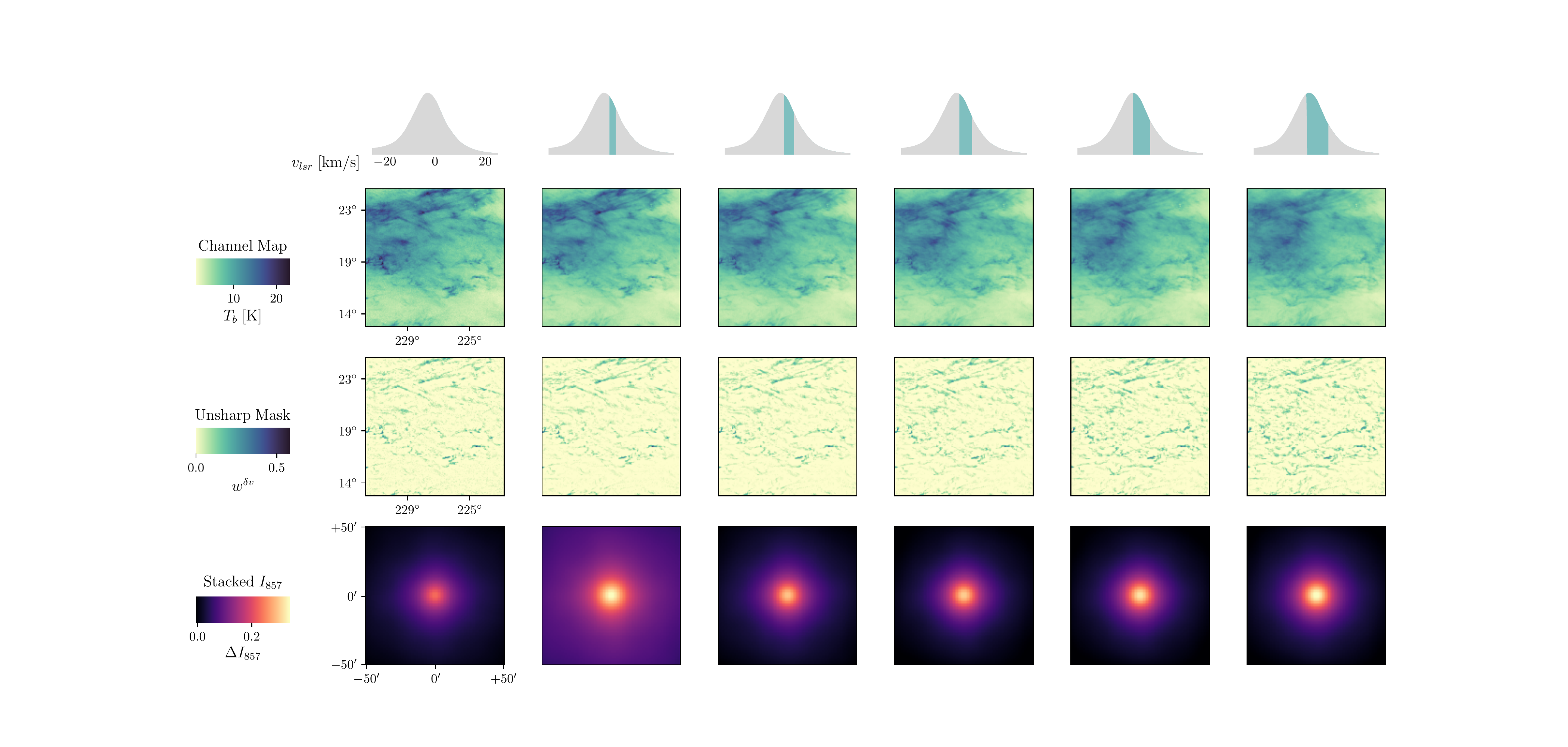}
\caption{Data and stacked data as a function of velocity channel width. Top panel: small cutout of channel map brightness temperature [K]. Middle panel: small cutout of unsharp mask. Unsharp mask intensity $w^{\delta v}$ is normalized such that the maximum value over the whole sky considered in this work is 1. Third panel: \textit{Planck} $857$ GHz intensity stacked on the nonzero pixels in the unsharp masked channel map data. Stacked FIR emission is plotted as $\Delta I_{857}$ [MJy/sr], the difference between the stacked values and the median value at the edge of the stack ($50'$ from the center pixel of the stack). Data are plotted on the same intensity scale with a common color bar. For all velocity channel widths there is a clear excess of $I_{857}$ emission at the center of the stacked data.}\label{fig:postagestamps}
\end{figure*}

\subsection{Thin-channel \hi intensity structures are not velocity caustics}\label{sec:USMnotcaustics}

In Section \ref{sec:RHTnotcaustics} we showed that elongated linear structures in thin-channel \hi intensity images correspond to dust-bearing density structures in the ISM. We made this measurement using the RHT, but other methods have been used to study the intensity structure of thin-channel \hi observations, such as the spatial gradient \citep[e.g.][]{LazarianYuen:2018}. Could it be that the structures that are highlighted by the RHT are special in terms of their physical nature -- that is, are \hi intensity features with strong RHT signatures more likely than other features to be real density structures? 

We test this by applying a different algorithm to the thin-channel \hi intensity data: the unsharp mask (USM). The USM is a high-pass filter, an operation that filters out low-frequency structure. Many popular algorithms for quantifying ISM spatial structure share this high-pass filter property, including the spatial gradient \citep{Gaensler:2012ix, PlanckXXXV:2016} and the Hessian \citep[e.g.][]{PlanckXXXII:2016}. The USM is used on its own to study ISM structure \citep[e.g.][]{Lee:2009, Kalberla:2016}, and is also the first step in the RHT. Applying the USM to the \hi channel map data highlights the small-scale intensity structure. 

We repeat the experiment of Section \ref{sec:RHTnotcaustics}, but stack the FIR and \nHI data on the USM of thin-channel \hi intensity rather than the RHT output.  
We find a similar result to the one summarized in Figure \ref{fig:stack_857_NHI_DGR}: there is a clear FIR and \nHI excess at the location of the stack. In Figure \ref{fig:postagestamps} we show small regions of \hi velocity channel data and their corresponding USM intensity. The \hi channel maps shown are all centered at $v_{lsr} \sim 0.4$ \kmsa, but vary in channel width from $0.18$ \kms to $7.9$ \kmsa. Beneath each data snapshot in Figure \ref{fig:postagestamps} we show the corresponding stack of $I_{857}$ on the USM over the region described in Section \ref{sec:data}. Each shows a clear excess of FIR emission at the center of the stack.
Thin-channel \hi intensity structures are not velocity caustics.

In the picture set forth by \citet{LP2000}, the degree to which the intensity structure is an imprint of the turbulent velocity field depends on the velocity channel width. The claim made by \citet{LazarianYuen:2018}, that intensity structures in channels of width $\delta v \sim 3$ \kms are velocity caustics, is already ruled out. Indeed, the correlation with the FIR shown in Figure \ref{fig:postagestamps} reveals that the structures in \hi channel maps are correlated with the density field down to the narrowest measured channel map width. 

\begin{figure*}
\centering
\includegraphics[width=\textwidth]{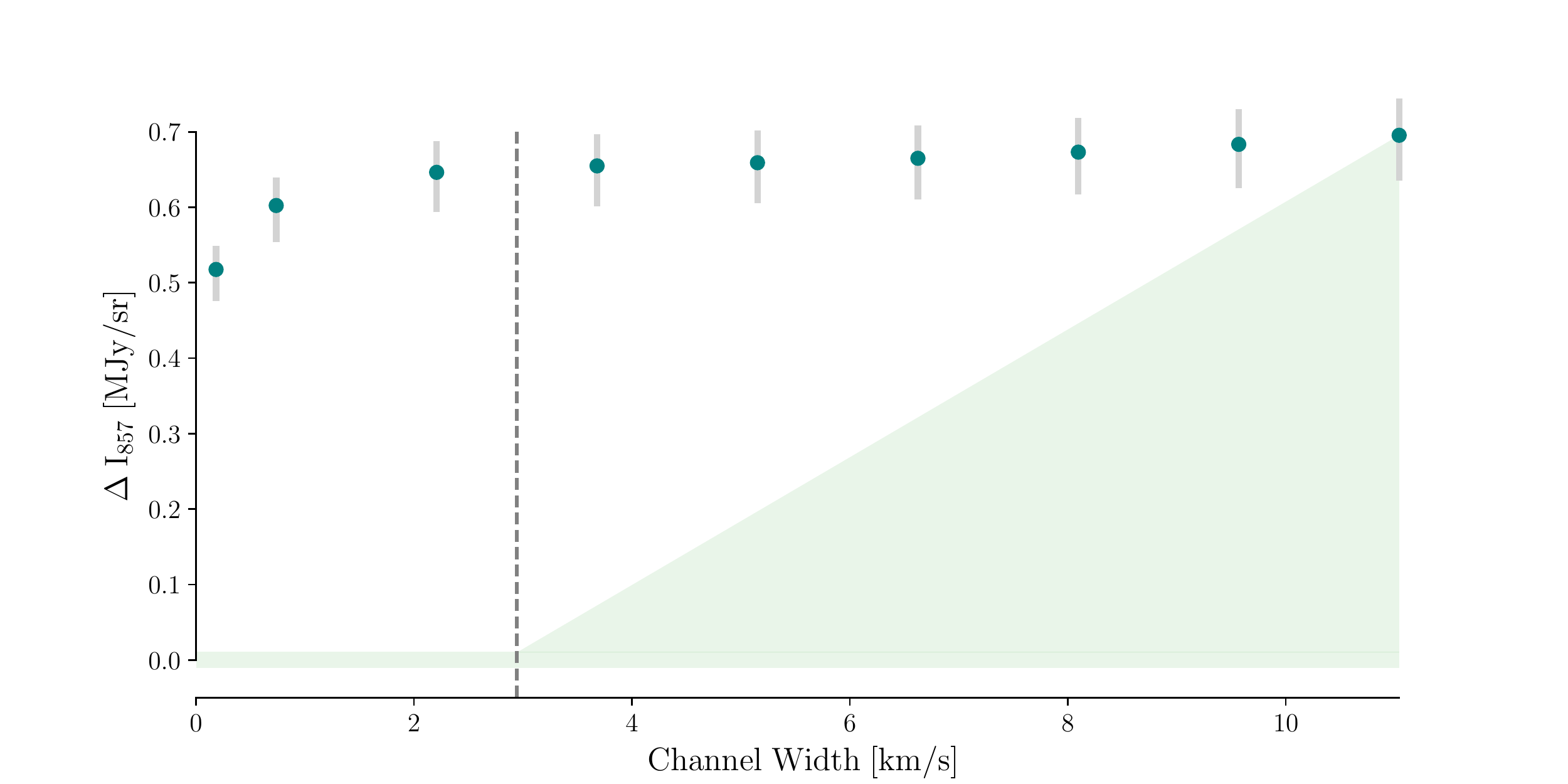}
\caption{The difference of the mean 857 GHz intensity weighted by the USM of \hi channel map data and the unweighted 857 GHz mean ($\Delta I_{857}$, Equation \ref{eq:deltaI857}). $\Delta I_{857}$ is plotted as a function of the velocity channel width of the \hi channel map to which the USM was applied. All velocity channels are centered at $v=0.4$ \kmsa. The pale green wedge is representative of the expected decline in $\Delta I_{857}$ if the \hi intensity features are velocity caustics. If the \hi intensity structures in the narrowest channels are purely or predominantly velocity caustics, the $w^{\delta v}$ in Equation \ref{eq:meanweightedI857} are uncorrelated with the underlying density field, and $\Delta I_{857} \rightarrow 0$. Instead, the data show a significant FIR excess associated with small-scale \hi intensity structures across all velocity channel widths. Error bars represent the $1\sigma$ distribution of $\Delta I_{857}$ measurements for block-bootstrapped data. Gray dashed line represents ${\delta v} = 2.94$ \kmsa, the velocity channel width used in \citet{Clark:2014, Clark:2015, LazarianYuen:2018}, and \citet{LazarianYuenHo:2018}.  }\label{fig:FIRwidth}
\end{figure*}

In Figure \ref{fig:FIRwidth} we show the results of an experiment to test the \citet{LP2000} prediction that narrower \hi velocity channels will be more dominated by velocity caustics. We compute the mean of the \textit{Planck} 857 GHz data, $\bar{I}_{857}$. This mean is computed over the region of diffuse, high Galactic latitude sky described in Section \ref{sec:data}. We then compute the weighted mean of the $I_{857}$ data, where the weights are the USM of the \hi intensity channel data. That is, we compute 

\beq\label{eq:meanweightedI857}
\bar{I}^{w}_{857} = \frac {\sum_{i=1}^{n} I_{i} w^{\delta v}_i} {\sum_{i=1}^{n} w^{\delta v}_i},
\eeq

where $I_i$ is $I_{857}$ in a given pixel, and $w_{\delta v}$ is the USM of an \hi intensity channel with velocity width $\delta v$.

We then compute

\beq\label{eq:deltaI857}
\Delta I_{857} = \bar{I}^{w}_{857} - \bar{I}_{857}.
\eeq

The quantity $\Delta I_{857}$ represents the average FIR enhancement correlated with small-scale \hi intensity channel structure. If the \hi intensity structure is caused entirely by velocity caustics, $w^{\delta v}$ will be uncorrelated with the density field, and $\Delta I_{857}$ will equal 0. 

\begin{figure*}
\centering
\includegraphics[width=0.9\textwidth, trim=0cm 0cm 0cm 0cm]{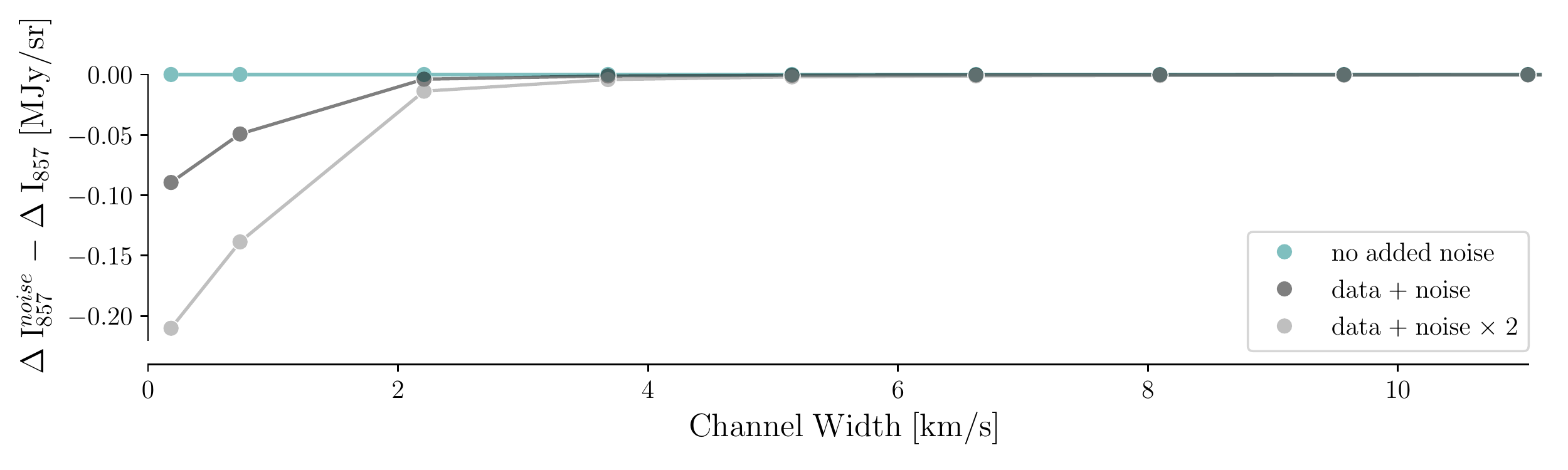}
\caption{The effect of additional noise on the $\Delta I_{857}$ measurement. We plot $\Delta I^{noise}_{857} - \Delta I_{857}$, the relative value between $\Delta I_{857}$ as shown in Figure \ref{fig:FIRwidth} and $\Delta I^{noise}_{857}$, or Equation \ref{eq:deltaI857} with additional estimated noise added to each \hi velocity channel. The noise contribution is estimated from an \hi intensity slice at $v_{lsr} = 400$ \kms as described in the text. This experiment demonstrates that the likely origin of the decrease in the value of $\Delta I_{857}$ in Figure \ref{fig:FIRwidth} is noise in narrow \hi velocity channels.
}\label{fig:appendix_FIR_noise}
\end{figure*}

As velocity caustics are expected to contribute more strongly to the \hi emission in narrower velocity channels, the caustics picture predicts a decrease of $\Delta I_{857}$ as $\delta v$ decreases. This caustics expectation is illustrated in Figure \ref{fig:FIRwidth} by a pale green wedge. While the slope and amplitude of the expected decline depends on details of the ISM density and velocity fields, the predicted trend is toward a more prominent contribution of the turbulent velocity field to the \hi intensity in narrower velocity channels. The gray dashed vertical line in Figure \ref{fig:FIRwidth} indicates the velocity channel width used in the \citet{Clark:2014,Clark:2015} papers and in the gradient-based analyses of \citet{LazarianYuen:2018} and \citet{LazarianYuenHo:2018}, to reflect the claim in the latter papers that structures in \hi intensity at these channel widths are velocity caustics. 

We calculate $\Delta I_{857}$ down to the narrowest $\delta v$ possible: the GALFA-\hi velocity resolution of $0.18$ \kmsa, well below the full width at half maximum (FWHM) linewidth of the coldest known \hi structures \citep{Peek:2011if}. Even with $\delta v=0.18$ \kmsa, $\Delta I_{857}$ is significantly inconsistent with $0$. This is a qualitatively identical result to the stacking experiment shown in Figure \ref{fig:postagestamps}: there is a significant $I_{857}$ GHz excess at the location of the USM structures for all $\delta v$. We find the same trend if we instead measure $\Delta$\nhi, the average \nHI enhancement correlated with small-scale \hi channel map structure, computed in analogy to Equations \ref{eq:meanweightedI857} and \ref{eq:deltaI857}. The channel maps used in this analysis are all centered at $v=0.4$ \kmsa, but the measured trend is qualitatively unchanged if we instead center the channel maps near the peak of the \hi line integrated over our region, $v \sim -3.3$ \kmsa.

To estimate the variance in $\Delta I_{857}$, we block bootstrap the data by dividing the selected sky area into 40 regions of roughly equal numbers of pixels. We sample these regions with replacement $10^7$ times and compute the $16^{\mathrm{th}}$ and $84^{\mathrm{th}}$ percentile ($1\sigma$) $\Delta I_{857}$ values from this distribution. These are visualized as the light gray error bars in Figure \ref{fig:FIRwidth}. As error bars, these are conservative and covariant, because they represent the sky variation rather than a statistical error.

Small-scale noise and telescope scan artifacts, more prominent in narrow velocity channels, will be uncorrelated with $I_{857}$ and will drive $\Delta I_{857}$ toward 0. To measure the effect of noise as a function of velocity channel width, we repeat the experiment shown in Figure \ref{fig:FIRwidth}, but prior to taking the USM of each velocity channel we add a ``pure noise" component. This noise map is the $\delta v = 0.72$ \kms GALFA-\hi emission centered at $v_{lsr} = 400$ \kmsa, multiplied by a factor $1/\sqrt{N}$, where $N = \delta_v/0.72$. We repeat the same experiment multiplying the noise component by $2$. The results are shown in Figure \ref{fig:appendix_FIR_noise}. The expected noise contribution has the same shape as the slight downturn in $\Delta I_{857}$ for the narrowest velocity channels in Figure \ref{fig:FIRwidth}. The amplitude of the slight downturn is likewise consistent with being solely caused by noise. \hi channel map noise similarly affects the stacked FIR emission in the narrowest velocity channels of Figure \ref{fig:postagestamps}.

\begin{figure*}
\centering
\includegraphics[width=\textwidth]{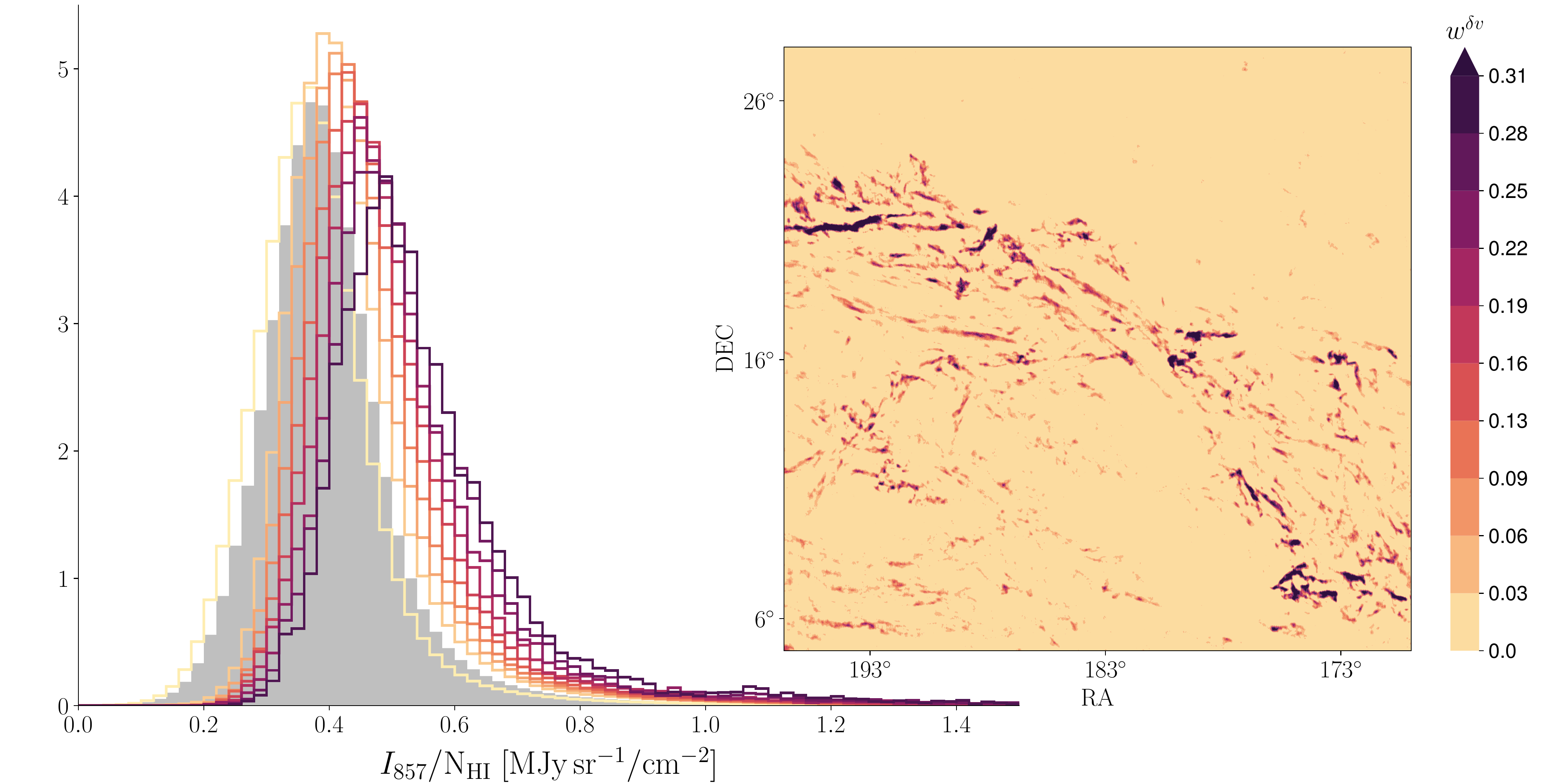}
\caption{Histogram of $I_{857}$/\nHI as a function of USM intensity. We compute this ratio for regions of the GALFA-\hi sky with $\left|b\right|>30^\circ$ and \nHI $< 8 \times 10^{20}$ cm$^{-2}$, excluding pixels in the HFI point source mask. In gray we plot this histogram for the full region of sky. We then partition the sky based on the intensity of the USM ($w^{\delta v}$) of an \hi velocity channel of velocity width $\delta v=2.2$ \kmsa. We divide the sky into 10 regions in equally spaced $w^{\delta v}$ bins that begin at 0 and end at the $99^{th}$ percentile of USM intensity. The $I_{857}$/\nHI histogram for each of these $w^{\delta v}$ bins is plotted on the left. For illustration, the USM of a small region of the \hi channel map is shown in the inset. Filled contours of the same $w^{\delta v}$ bins are shown in the same colors as the histograms. }\label{fig:FIR_hist_by_USM}
\end{figure*}

Neither $\Delta I_{857}=0$ nor a decline in $\Delta I_{857}$ with decreasing velocity channel width is consistent with the data. The FIR associated with small-scale channel map structure is significantly higher than the sky average for all velocity channel widths, and remarkably flat as a function of channel width. Only for the narrowest channel widths, $\delta v< 1$\kmsa, do we measure even a marginal decrease in $\Delta I_{857}$. This $\sim15\%$ decrement in $\Delta I_{857}$ is consistent with noise in the \hi intensity maps. Evidently, not only are thin-channel \hi intensity structures not dominated by velocity caustics, but there is no measurable contribution to the \hi intensity from velocity caustics at all.  

\subsection{Thin-channel \hi intensity structures are preferentially CNM}\label{sec:CNM}

We have so far established that high-spatial frequency structures in narrow \hi velocity channels are density structures, not velocity caustics. Here we investigate the physical nature of these density structures, and suggest a physical picture of the ISM that is consistent with these observations. 

We examine the distribution of FIR/\nHI for structures at different spatial scales. Figure \ref{fig:FIR_hist_by_USM} shows the (normalized) histogram of $I_{857}$/\nHI as a function of the intensity of the USM ($w^{\delta v}$) of an \hi intensity channel with width $\delta v=2.2$ \kmsa. Higher values in the USM correspond to channel map features that are higher contrast with respect to the local low-frequency spatial structure. The gray histogram shows the $I_{857}$/\nHI for the entire region of sky considered. The histogram of $I_{857}$/\nHI clearly shifts toward higher values as sky pixels with higher USM values are selected. The high spatial frequency structures that the USM highlights have preferentially higher $I_{857}$/\nHI than the sky as a whole. The inset in Figure \ref{fig:FIR_hist_by_USM} shows the USM of a small region of sky. The structures highlighted tend to be elongated, a fact discussed further below. 

The FIR/\nHI associated with the small-scale structures is consistently higher than the sky-averaged value. This suggests that high spatial frequency \hi intensity structures are physically distinct from the surrounding medium, beyond simply having higher density. There are three physical effects that are expected to raise the FIR/\nHI ratio. These are \citep[e.g.][]{Ysard:2015,Nguyen:2018}:

\begin{enumerate}
\item Optically thick \hi emission, which will lower the \hi column without affecting the associated FIR emission.
\item Spatially correlated molecular hydrogen ({H}$_2$), which depletes the population of atomic hydrogen, thereby lowering \nhi.
\item Increased dust emissivity or dust-to-gas ratio associated with dense gas, which will raise the FIR emission relative to \nhi. 
\end{enumerate}

These three possible contributing effects are notoriously difficult to disentangle \citep{Burstein:1978, Boulanger:1996, Lenz+Hensley+Dore_2017}. Nevertheless, all three of these physical situations are associated with the cold phase of the ISM (referring both to CNM and to the diffuse molecular phase). 

The enhanced FIR/\nHI cannot be due to noise bias from weighting the data by the USM of an \hi channel map. The primary effect of radiometer noise in a narrow velocity channel will be to add spurious power both in that narrow channel map and in the \nHI map. This correlated noise will correspond to an enhancement in \nHI but not in the dust tracer. This bias will thus tend to suppress, rather than enhance, the FIR/\nHI ratio, and so noise bias cannot account for our result.

We therefore expect that the observed enhancement of the FIR/\nHI ratio correlated with small-scale \hi intensity structure is because these channel map structures are preferentially CNM. Although this is, to our knowledge, the first measurement showing that high spatial frequency \hi channel map structures are associated with higher FIR/\nhi, the picture that CNM structures are preferentially small-scale is consistent with the current understanding of the ISM phase distribution. Indeed, while CNM accounts for roughly half of the \hi mass fraction, the CNM volume filling fraction is of order 1\% of the WNM volume \citep{Heiles:2003dh, Murray:2018}. This alone implies that the CNM is preferentially distributed into small-scale structures: the CNM is clumpy compared to the extended, diffuse WNM. 

\section{Discussion}\label{sec:discussion}

We have shown that the small-scale structure in narrow \hi channel maps is predominantly cold phase material, and the preferential elongation of these density structures in the direction of the ambient magnetic field is responsible for the measured alignment between \hi intensity structure and the plane-of-sky magnetic field orientation. Other observations bolster the picture that the magnetically aligned \hi intensity structures are CNM. Filamentary structures in narrow \hi channel maps have FWHM linewidths consistent with CNM temperatures \citep{Clark:2014, Kalberla:2016}. Most strikingly, observations of \hi self-absorption in the Riegel-Crutcher cloud show that the CNM is organized into slender, linear structures that are very well aligned with the local magnetic field \citep{McClureGriffiths:2006wx}. Our results suggest that, if \hi self-absorption imaging were possible for more of the sky, such a cloud structure would not be uncommon. Indeed, taken together with the results of \citet{Clark:2014,Clark:2015}, we suggest that the CNM is \textit{generically} organized not into spherical clumps, but into magnetically aligned filamentary structures. The prediction of this work, that high spatial frequency \hi intensity structures are preferentially associated with cold gas, should be tested with absorption line measurements of the \hi spin temperature. The Square Kilometer Array (SKA) and the Australian SKA Pathfinder (ASKAP) will allow this picture to be tested in many different Galactic environments \citep{2015aska.confE.130M, Dickey:2012wf}.

How do the magnetically aligned CNM features in \hi channel maps compare to magnetically aligned FIR intensity structures? \citet{PlanckXXXV:2016} used the spatial gradient of the column density to characterize the local orientation of dust structures. The authors reported an alignment between the dust intensity structures and the magnetic field for low column densities, and quantified the correlation statistically using the Histogram of Relative Orientations \citep[][]{Soler:2013dh}. \citet{PlanckXXXII:2016} similarly find that dust structures are well aligned with the projected magnetic field at lower column densities. \citet{LazarianYuen:2018} refer to the spatial gradient of column density maps as ``intensity gradients", and argue that they do not trace the magnetic field as well as gradients of velocity channels. The authors find that gradients of thin channel maps (``velocity channel gradients") are better correlated with the local magnetic field orientation than gradients of either the column density map (``intensity gradients") or of the first moment of the velocity (``velocity centroid gradients") \citep{YuenLazarian:2017, GonzalezCasanova:2017}. \citet{LazarianYuen:2018} (and references within) argue that this is because thin channel maps trace the velocity-magnetic field alignment, and so the higher-fidelity alignment of narrow velocity channels confirms the theoretical expectation that the density-magnetic field correlation is weak compared to the velocity-magnetic field correlation \citep[e.g.][]{ChoLazarian:2003, Passot:2003}.
The present work overturns that interpretation.

Why, then, are the \hi channel map structures better aligned with the magnetic field than structures in \nhi?
In their analysis of the relative orientation of the magnetic field and ridges in the FIR emission, \citet{PlanckXXXII:2016} estimate volume densities for the dust filaments that are consistent with CNM or diffuse molecular gas, although they primarily analyze higher column density features (median $\mathrm{N}_\mathrm{H} = 1.2\times 10^{21}$ cm$^{-2}$) than considered here. \citet{PlanckXXXII:2016} find that dust structures are better aligned with the projected magnetic field at lower column densities. To the extent that the magnetic alignment is stronger for structures in the \hi channel maps, it may be because diffuse CNM structures have a higher contrast relative to the surrounding medium when the \hi intensity is integrated over a narrow velocity range. Projection effects and line-of-sight effects can cause the observed correlation between the plane-of-sky magnetic field and intensity structures to vary from the true relative orientation between the three-dimensional magnetic field and density structures. 21-cm observations may thus be a useful tracer of the three-dimensional magnetic field, because line-of-sight effects can be probed by analyzing multiple channel maps along the line of sight \citep{Clark:2018}.

This work calls into question the interpretation of other analyses that are premised on a significant contribution of the turbulent velocity field to structures in narrow \hi velocity channels. Based on this expectation, \citet{LP2000} proposed that the three-dimensional velocity spectral index can be extracted from spectral observations by measuring the power spectral index of channel maps of varying $\delta v$ \citep[see also][]{Pogosyan:2005}. \citet{LP2000} predict a steepening of the power spectrum slope with increasing $\delta v$, as a result of a decreasing contribution of the velocity field to the intensity fluctuations. One observational test of this picture in \hi is given in \citet{Pingel:2013}, where indeed the slope of the two-dimensional spatial power spectrum (SPS) steepens with increasing velocity slice thickness. Our work suggests that an alternative explanation for this result should be considered: that the intensity contribution of CNM structures increases with decreasing $\delta v$. If the CNM has a shallower power spectrum than the warmer-phase gas, this would also reproduce the qualitative trend of steeper SPS for larger $\delta v$. The preferential distribution of CNM into small-scale structures is consistent with a shallower SPS for the CNM. Other observational analyses that interpret shallower channel map spectral indices as a signature of velocity fluctuations should be reevaluated in light of this work \citep[e.g.][]{Dickey:2001}, and should include a careful treatment of the noise \citep[e.g.][]{Blagrave:2017}.

\section{Conclusions}\label{sec:conclusions}

We draw a number of conclusions about the physical nature of \hi intensity maps, the structure of the multiphase ISM, and the interpretation of synthetic channel maps. These are summarized as follows:

\begin{itemize}
\item Small-scale structures seen in narrow velocity channels are dust-bearing density structures. A description of \hi intensity structures as being entirely or predominantly velocity caustics is incorrect.

\item Techniques based on the gradient of intensity structures in narrow spectral channels successfully trace the orientation of the interstellar magnetic field because they measure the orientation of density structures that are preferentially aligned with the magnetic field, not because they measure an imprint of the turbulent velocity field. 

\item Small-scale \hi channel map structures have higher FIR/\nHI than their surroundings, which implies that they are associated with colder, denser phases of the ISM.

\item The data presented are consistent with a picture of the neutral ISM in which the colder-phase gas is measurably distributed into small-scale, anisotropic structures, preferentially aligned with the magnetic field. The WNM, by contrast, is more extended and isotropic. 

\item A number of numerical studies predict a significant influence of the turbulent velocity field on the structure of \hi channel maps, in conflict with the data. We discuss the origin of some misleading theoretical expectations. Neglecting thermal broadening is unphysical and artificially inflates the intensity contribution from velocity caustics. Thermal broadening in subsonic \hi smooths out velocity-induced intensity fluctuations and correlates channel maps with the density field. Supersonic motions dynamically correlate the three-dimensional density and velocity fields. We demonstrate that these effects correlate synthetic \hi channel maps near the peak of the line with the synthetic column density map regardless of sonic Mach number.

\item A shallower spatial power spectrum slope measured in narrower velocity channels is qualitatively consistent with a higher contribution to the intensity in narrow channels from CNM gas. Previous work has attributed such a change in the spatial power spectrum to velocity fluctuations, and used this to derive the slope of the turbulent velocity power spectrum in the gas. This work calls those analyses into question.

\end{itemize}

This work calls for a significant reassessment of many observational and theoretical studies of turbulence in \HI.

\software{astropy \citep{Astropy:2013, Astropy:2018}, cmocean \citep{cmocean:2016}, matplotlib \citep{Matplotlib:2007}, numpy \citep{Oliphant:2015:GN:2886196}}

\acknowledgments
We thank Fran\c cois Boulanger for many enlightening discussions over the course of this investigation, and for reading an early version of this manuscript. We thank Claire Murray and Ludovic Montier for illuminating conversations on this work and related topics.
We thank C\'edric Colling and Patrick Hennebelle for the numerical simulations. This work took place in part under the program Milky-Way-Gaia of the PSI2 project funded by the IDEX Paris-Saclay, ANR-11-IDEX-0003-02. We thank the other participants of that program for stimulating discussion. We thank the Flatiron Institute and Chang-Goo Kim for hospitality at the Center for Computational Astrophysics, where some of this analysis took place. We thank the anonymous referee for a thoughtful and prompt review.
S.E.C. is supported by NASA through Hubble Fellowship grant \#HST-HF2-51389.001-A awarded by the Space Telescope Science Institute, which is operated by the Association of Universities for Research in Astronomy, Inc., for NASA, under contract NAS5-26555. This publication utilizes data from Galactic ALFA \hi (GALFA-\HI) survey data set obtained with the Arecibo L-band Feed Array (ALFA) on the Arecibo 305m telescope. The Arecibo Observatory is operated by SRI International under a cooperative agreement with the National Science Foundation (AST-1100968), and in alliance with Ana G. M\'endez-Universidad Metropolitana, and the Universities Space Research Association. The GALFA-\hi surveys have been funded by the NSF through grants to Columbia University, the University of Wisconsin, and the University of California.

\bibliography{refs}

\end{document}